# Stable halogen 2D Materials: the case of iodine and astatine

Xinyue Zhang[a], Yu Liu[b] and Qingsong Huang*[a]

**Abstract:** Two-dimensional (2D) materials have wide applications towards electronic devices, energy storages, and catalysis, et al. So far, most of the pure element 2D materials are composed of group IIIA,IVA, and VA elements. Beyond the scope, the orbit hybrid configuration becomes a key fact to influence 2D structure stably. Here we show a $sp^2d^3$ hybridization in the outmost electrons with O-shell for Iodine and P-shell for astatine element, builds up triangle configuration (β-type) to form 2D structures-β-iodiene and β-astatiene. Each atom is connected by σ bonds, and surrounded by 6 atoms. The π bonds become possible, and band gap approaches zero because of interaction of unpaired single electron to each atom, depending on reducing bond length. By applying compression strain or spin orbit coupling (SOC), the Dirac points or topological nontrivial points can be available in the β-iodiene and β-astatiene. Our discovery has paved a new way to construction of 2D materials.

## Introduction

Two-dimensional materials have become a star in boosting development of new energy[1,2], catalysis[3,4], and electronic devices[5] etc. So far, single-elemental 2D materials have all been built up mainly by the elements of main group IIIA, IVA, and VA[6] and so on. Typically, graphene and some graphene-like structures have been found by now, such as borophene [7,8], indiene (group IIIA)[9], silicene [10-14], germanene[15-18],stanene[19-22] (group IVA),black phosphorene(BP)[23-26], bismuthene[27-29], and arsenene (group VA)[30-32], etc. Some 2D materials, e.g. graphene, the unpaired p-wave electron can form weak π bonds, inducing linear-energy dispersion and mass-less Dirac-Fermions[33]. Whereas VA group elements, such as BP, arsenic, antimony, and bismuth, form two-dimensional configurations by sp3-hybridized orbit, and their topological and nontrivial point or Dirac points arise from appearance of band inversion[34-37], instead of formation of π bonds[38]. The π bond and linear-energy dispersion locks the survival of Dirac-points, implying the existence of massless Dirac-Fermions [39-40]. By far, the 2D materials from group IIIA-VA are generally recognized as forming sp2 or sp3 hybrid orbits, exhibiting covalent σ bond strong enough to sustain the 2D configurations[41].

[a] Prof Qingsong Huang, Ms Xinyue Zhang
School of chemical engineering
Sichuan University
Chengdu, 610065, P.R.(China)
E-mail: qshuang@scu.edu.cn
[b] Dr.Yu Liu
Center for Quantum Devices
University of copenhagen & Microsoft Quantum Materials Lab
Copenhagen
Denmark

Supporting information for this article is given via a link at the end of the document

However, construction of 2D materials by elements beyond IIIA, IVA and VA main group, e.g., VIIA remains challenging by now. Usually, elements from VIIA are difficult to form periodical configuration because of violation of octet rule, weak covalent bonding, and hybridization rule, involving *d* band in same shell of outmost electrons (high energy difference between *p* wave orbits and d wave orbits).

Since the low difference between *p* wave orbits and *d* wave orbits of I and At atoms, the *spd* hybridization can be realized, whereas the covalent bonding in elemental 2D materials seems weaker than those in group IIIA-VA. Our previous work adopted the DFT simulation to verify the existence of a planar iodine structure with square cell - iodiene [42].

Here we present 2D materials derived from element iodine or astatine in group VIIA. The 2D structure is arising from triangle configuration with named as β type configuration of iodiene or astatiene, illustrating the difference from the square cell.

The β-iodiene and β-astatiene can be constructed by σ bonds as strong as that arising from VA elements. Based on skeleton of $sp^2d^3$ hybridized σ band, π band in high symmetry points is difficult to form since the long bond length (LBL), despite the unpaired electron is available for each atom [43-45]. Formation of Dirac point depends on band inversion [46], arising from bond length reductions to form π bond [44], or perturbation of Anderson potential barrier [47]. The phonon spectrum calculation shows that iodine and astatine can form stable two-dimensional materials（Fig. 1a,b）. Provided absence of strain, the LBL is too far to form π band. When compressing strain is applied, the LBL becomes smaller than before, and a Dirac point comes into being finally.

## Results and Discussion

Iodine or astatine element can form planar 2D configurations respectively (Fig. 1a-b), building up β-iodiene and β-astatiene structures. In addition, the triangular cell configuration, instead of hexagonal (Fig. S7), is indeed stable, suggesting the hexagonal cell and bucking configurations are all unstable after undergoing the geometrical optimization (Fig. S17- Fig. S20). The *P6/MMM* spatial point group is sustained by $sp^2d^3$ orbits, demonstrating a space lattice of inserting an atom into the center of hexagonal two-dimensional crystal cell. Planar-2D β-iodiene and β-astatiene structures are stable, because of absence of imaginary frequency in phonon spectrum (Fig. 1c-d). While the triangular cell with bucking configuration prefers to underlying planar structure after undergoing the geometrical optimization (Fig. S17, Fig. S19). Since there is totally one atom in their primitive cell, only acoustic branches are available in the phonon spectra.

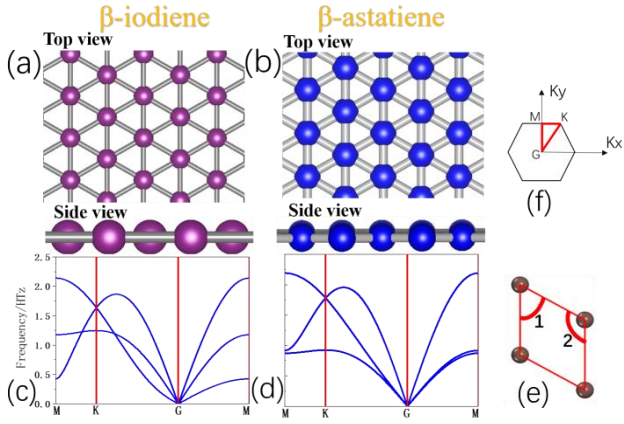

**Figure 1** (Color online) The typical 2D configuration and stability of β-iodiene and β-astatiene. Ball and stick model of (a): β-iodiene, (b): β-astatiene. Phonon spectra of (c): β-iodiene and (d): β-astatiene. (e): Schematic diagram of the primary cell of β-iodiene and β-astatiene structures. (f): Brillouin Zone.

Moreover, the positive $E_b$ (Binding Energy) of β-iodiene and β-astatiene also provide evidence that they can be stable (Table 1). The lattice constants of β-iodiene and β-astatiene structures are 3.37 Å and 3.51 Å respectively, which are larger than graphene and boron nitride[48]. β-iodiene and β-astatiene are broad band-gap materials with gaps of 2.85 eV and 2.25 eV respectively (Fig. 2). Similar wide-band gap two-dimensional materials are arsenene (2.49 eV) and antimonene (2.28 eV)[31].

**Table 1.** The properties of β-astatiene and β-iodiene: total energy ($E_{total}$), binding energy ($E_b$), lattice constant (l), bond length(b), bond angle (θ), bandgap ($E_{gap}$)

| type | $E_{total}$/eV | $E_b$/eV[a] | l/ Å | b/ Å | $θ_1$[b] | $θ_2$[b] | $E_{gap}$/eV |
|---|---|---|---|---|---|---|---|
| β-iodiene | -1.21 | 1.14 | 3.37 | 3.37 | 60 | 120 | 2.85 |
| β-astatiene | -1.29 | 1.26 | 3.51 | 3.51 | 60 | 120 | 2.25 |

[a] Binding energy is calculated according to formula (1). [b] Bond angle $θ_1$, $θ_2$ refer to Fig. 1e. The bond length is the same as the lattice constant.

Here, no Dirac points is available in the β-iodiene and β-astatiene band structures (Fig.2a,c). Due to the LBL of β-astatiene (3.51 Å) and β-iodiene (3.37 Å), the unpaired single electron become localized, suggesting $P_Z$ orbitals cannot overlap in space, and π bond becomes transient and unsteady (Fig. 2a,c). The interlayer spacing is around 3.35 Å for graphite by Bernal stacking, and the interacting between the contiguous layers is just Van der Waals force instead of covalent (chemical) bonds. Here, the bond lengths of β-astatiene (3.51 Å) and β-iodiene (3.37 Å) are smaller than van der Waals diameters (4.30 Å)[49], suggesting the covalent is available, and strong enough to support the 2D structure, in comparison to phosphorene [19]. Whereas, their LBLs are much larger than that of graphene (1.42 Å) with survival of π bond, and Dirac points (Fig. 2b).

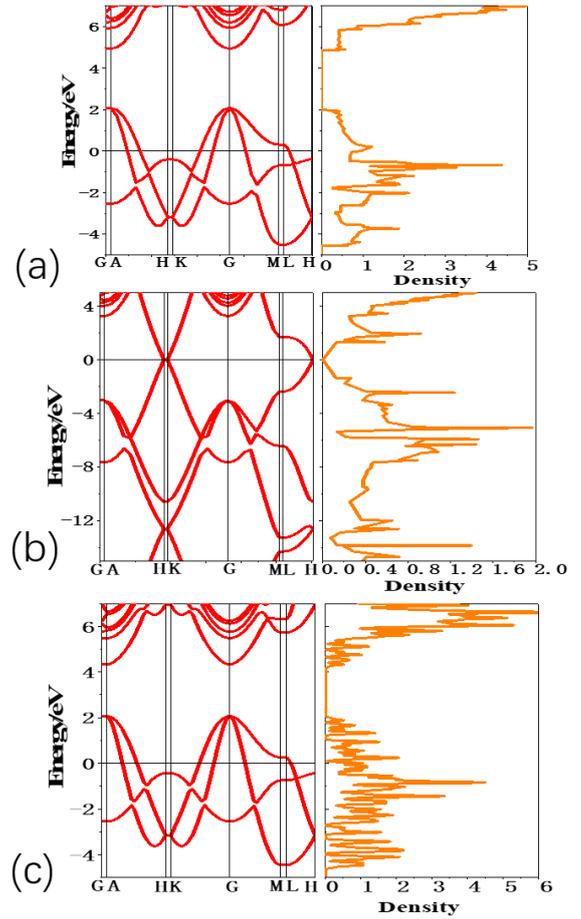

**Figure 2** (Color online) DFT simulation of band structure and DOS (density of state). (a): β-iodiene, (b): graphene, and (c): β-astatiene. At Gamma point, the β-iodiene and β-astatiene own direct band gap (a,c), whereas the π band at K point comes into being with constructing carbon atoms into graphene configuration (b), implying Dirac Fermions arises from the linear dispersion over band structure.

Although no Dirac point is available in the raw 2D materials of the β-iodiene and β-astatiene, their band structure can be further tuned by strain. When biaxial tensile strain is applied to β-iodiene, its band gap increases (Fig. 3a₁). When applying biaxial compression strain, the band gap of β-iodiene became reduced with the strain increasing (Fig. 3a₃-a₅). When the strain reaches 10.2% (Fig. 3a3), the band gap of β-iodiene was reduced to zero. After that, further increase of strain can induce the band inversion, leading to the formation of topological nontrivial points, one type of Dirac points[43], suggesting the β-iodiene is one possible topological semimetal system[50], or Quantum Spin Hall (QSH) system[51]. When the strain increases to 12.5%, a striking inversion band can be observed (Fig. 3a₄). Dirac points burst out near the high symmetry point of Gamma point in Brillouin zone (BZ), coming from the band inversion of the valance band and Conduct band. Even more Dirac points and further band inversion involving in more energy level can be observed if strain is increased to 15% (Fig. 3a₅).

β-astatiene undergoes a same process with compression strain increasing (Fig. 3b₁-b₅). When the compression strain arrives 9% (Fig. 3b₃), the band gap of β-astatiene was reduced to zero. After that, if the strain increases to 12.5%, Dirac points and topological nontrivial points burst out near the Gamma point in BZ, because of valence band inversion into conduction band (Fig. S1). Further band inversion can be observed when strain increases to 15% (Fig. 3b$_5$). The section under 3.6 eV is made in the β-astatiene 3$d$ band structure by 15% compression strain (Fig. 3c). The junction of valence band and conduction band is shown by red lines, showing a slightly concave hexagon, and valence band is shown by blue lines, showing an ellipse. The convex part of the valence band is embedded in the concave part of the conduction band.

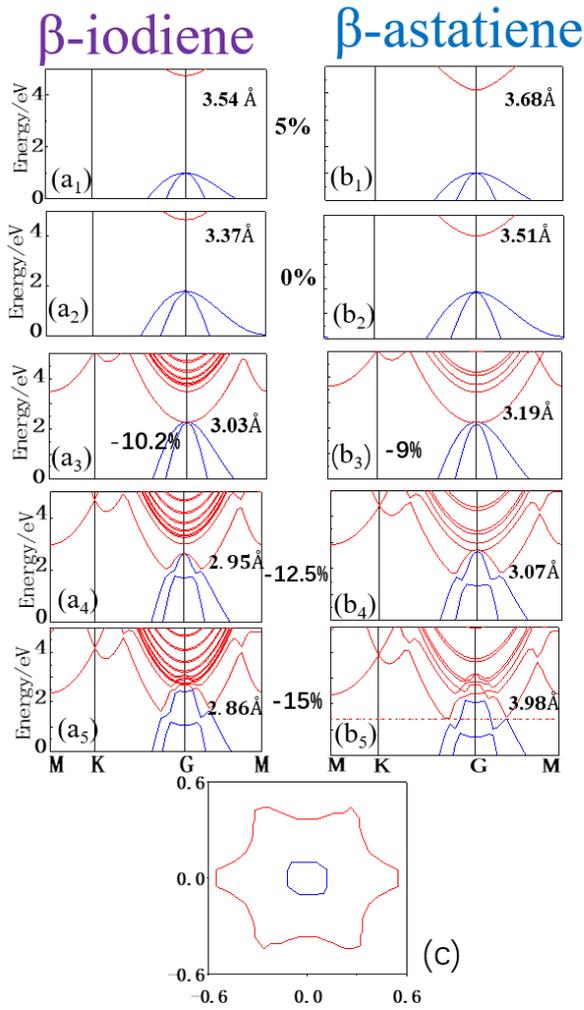

**Figure 3** (Color online) Band structures of iodiene and astatine under biaxial strain. Modification of β-iodiene band structure by applying different biaxial strains of (a1):5%, (a2):0%, (a3): -10.2%, (a4):-12.5% and (a5):-15%. While the band structure of β-astatiene changes with applying different strains of (b1):5%, (b2):0%, (b3):-9% (b4):-12.5%, (b5):-15%. A positive strain represents a tensile strain and negative strain represents the compression strain. (c): The cross section of the red dotted line in b5. The intersection of the conduction band and valence band obtains a slightly concave hexagon (shown in red), corresponding to the symmetry of p6/mmm space. The valence band is shown by the blue line.

Figure 4 illustrates the three-dimensional (3D) band structure of β-iodiene under biaxial strain. The HOMO and LUMO of β-iodiene are direct band gap at the Gamma point (Fig. 4a). As the compression strain increases, the HOMO and LUMO move closer to each other until the strain reaches 10.2%, where band gap approaches zero (Fig. 4b). Beyond the critical strain (10.2%), Dirac arcs (strain=12.5%) or topological nontrivial lines are available (Fig. 4e, Fig. S2 a5), arising from the inversion of the valence band and conduct band (Fig.4c). Furthermore, the strain 15% leads to Dirac circle lines or topological nontrivial lines are available, suggesting the properties of Dirac semimetal[50] (Fig. 4f, Fig.S2 b6).With the strain increasing, three dimensional band structure of β-astatiene will undergo a same history as that of β-iodiene (Fig. S2, Fig. S3). The β-iodiene 3D band structures of HOMO and LUMO have been disconnected to be illustrated respectively under the 15% biaxial compression strain (Fig. 4g-h). Here the strain is adopted to promote the HOMO band level and the band gap will be reduced simultaneously. Once the HOMO becomes contacted with LUMO, the further compressive strain will lead the bands, both conductance band and valance band, to reverse around the gamma point (Fig. 3a$_5$). The LUMO shows obvious band inversion under strain, but the LUMO not (Fig. S6 a$_1$-a$_2$, Fig. S6b$_1$-b$_2$).

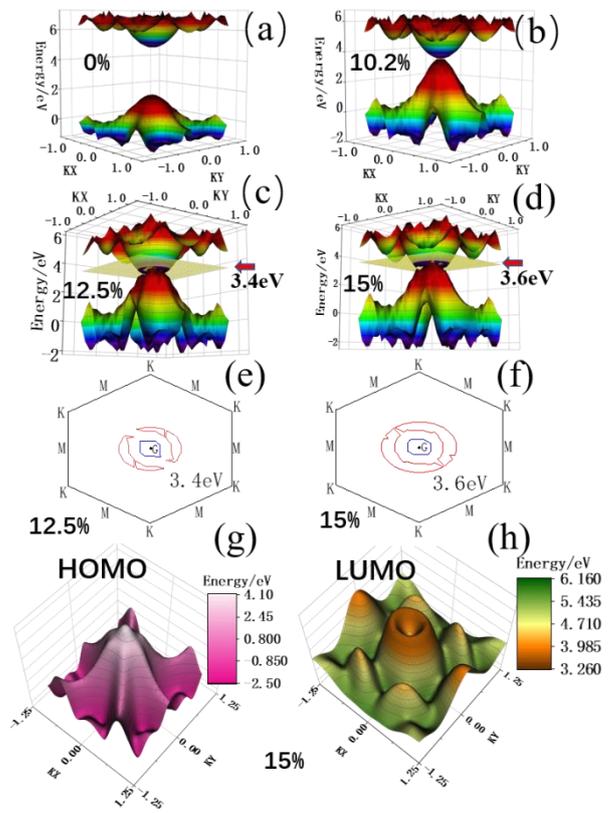

**Figure 4** (Color online) The β-iodiene 3D band structure at (a):0%, (b):10.2%, (c):12.5%, (d):15% biaxial strain. Cross section of β-iodiene band structure at (e)3.4 eV under 12.5% strain, (f) 3.6 eV under 15% strain. The red line represents the cross section of the conduction band and the blue line represents the valence band. The red arrows in c and d indicate the positions of cross section figure e and f respectively. (g):3D HOMO band structure of β-iodiene under 15% biaxial strains. (h):3D LUMO band structure of β-iodiene under 15% biaxial strains. The LUMO structure has been turned 180° to its back.

Under uniaxial strain, the structure symmetry will be damaged, and the band structure undergoes different evolution history from biaxial strain state (Fig. 5). Since high symmetry has lost, their properties turn to anisotropy. For β-iodiene, even 12.5% strain cannot reduce the band gap to zero (Fig. 5a, e), and the band structure has lost its symmetry near the gamma point (Fig. 5a, e). When uniaxial strain reaches 15%, the band inversion makes Dirac points or topological nontrivial points available (Fig. 5c, f). The LUMO and HOMO bands contact and develop into some Dirac arcs, suggesting a Dirac semi-metal state or topological semi-metal state is available (Fig. 5f). The damaged symmetry makes the inversion bands contact at partly between LUMO and HOMO (Fig. 5f). The β-astatiene will undergo same process with the increasing of uniaxial compression (Fig. 5b, d), except for the band inversion initiating much earlier than that in β-iodiene. The difference is band inversion in β-astatiene occurs much earlier than that in β-iodiene. In fact, under 12.5% strain, the band inversion can be observed in β-astatiene (Fig 5.b), implying the properties of Dirac semimetal or topological semimetal can be available. The 3D band structure evolution can be found more in detail referring to Fig. S4-Fig. S5. Although the uniaxial strain reduced the symmetry of 2D configuration, the band inversion remains happened in LUMO centering around the Gamma point (Fig 5.h, Fig. S6$c_1$-$c_2$, Fig. S6$d_1$-$d_2$).

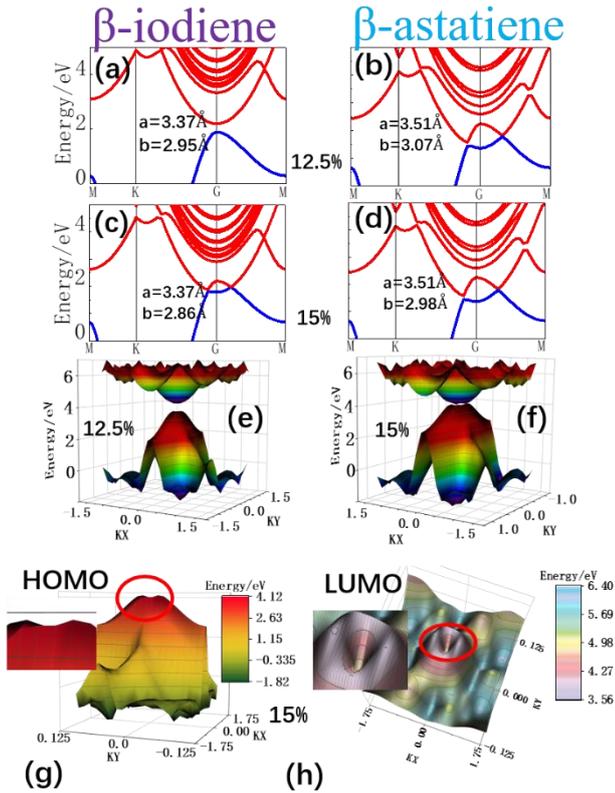

**Figure 5**(Color online) uniaxial strain band structures of β-iodiene and β-astatiene. Band structure under 12.5% of (a): β-iodiene, (b): β-astatiene, and under 15% (c): β-iodiene, (d): β-astatiene. 3D band structure under 12.5% of (e): β-iodiene, and under 15% (f): β-iodiene. (g):3D HOMO band structure of β-iodiene under 15% uniaxial strains. (h):3D LUMO band structure of β-iodiene under 15% uniaxial strain. The LUMO structure has been turned 180° to its back.

Hexagon β-iodiene shows a significant imaginary frequency in the phonon spectrum (Fig. S7 a,c), implying such 2D configuration is unstable. Its Electron localization function (ELF) section diagram shows a strong anti-localized [52] region in the centre of the hexagon structure (Fig. 6a). The connection bond energy looks low, implying a weak bonding state between the iodine atoms. In an unit cell, an iodine atom can be introduced to centre of the hexagon (Fig. 6b) and create six triangular periodic structure, removing imaginary frequency completely from this structure(Fig. 1c). Although the introduced additional atom in the centre cannot promote the bonding energy for single bond (Fig. 6a-b), the increasing bond density enhance the bonding level substantially (Fig. 6b). Since the values weak close to 0(blue) for vacuum (Fig.6 legend), moderate (green) for homogeneous electron gas, and strong close to 1(red) for perfect localization, the higher the ELF value of a considered point is, the higher the localization degree of electrons in this region is indicated.

The ELF diagram of Hexagon β-astatiene shows an unstable structure, mimicking that of β-iodiene (Fig.6c), because of imaginary frequency in phonon spectrum (Fig. S7 b, d). Whereas, the colour of the β-astatiene ELF diagram shows that bond between astatine atoms (Fig. 6c) is stronger than that surrounding iodine atom (Fig. 6b). For both structures, ELF around the atom is between 0.2 and 0.4, which is a weaker covalent bonding state than the normal one in group IIIA-VA [52].

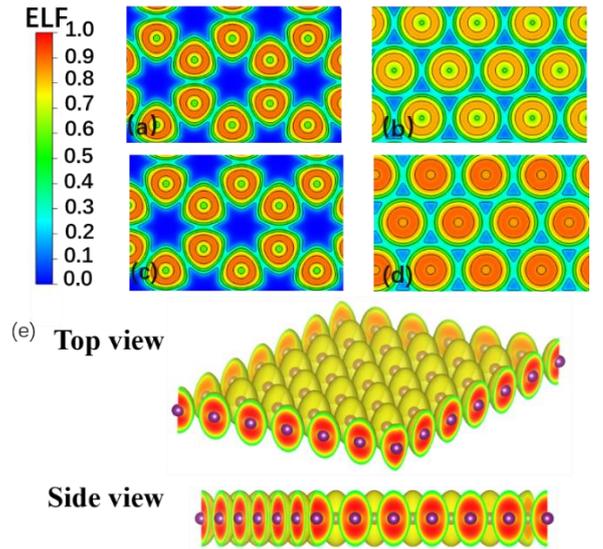

**Figure 6**(Color online) Electronic local functions (ELF) of β-iodiene and β-astatiene. (a): Hexagon β-iodiene. (b): β-iodiene. (c): Hexagon β-astatiene. (d): β-astatiene. Obviously unstable hexagon structures have larger electron vacuum regions than triangular structures. (e): Top and side view of β-iodiene electronic local functions of 7×7 super-cell.

In addition, the ELF can be illustrated in three dimensional images (Fig. 6e). The bond length is 3.37 Å for iodiene, and 3.51 Å for astatiene, which is much shorter than the Van deer Waals diameter [51], suggesting a covalent bond is available. Each atom has six nearest neighbors (Fig. 6e). Since each atom with unpaired single electron is connected by σ bonds with surrounding 6 atoms, the π bonds become possible, depending on the bond length.

Fig. 7 shows the Partial density of states (PDOS) of β-iodiene and β-astatiene near Fermi surface, demonstrating the $s$, $p_z$, $p_x$, $d_{xz}$, $d_{yz}$, $d_{zz}$ in β-astatiene and β-iodiene are participating in orbital hybridization, belonging to $sp^2d^3$ orbit hybrid within the O shell for β-iodiene and P-shell for β-astatiene. The contribution of each wave can consult the Fig. 7. The outmost electrons in β-iodiene (Fig. 7a-b) and β-astatiene (Fig. 7c-d) perform $sp^2d^3$ hybridization. In β-iodiene, 6 electrons will occupy 6 hybridization orbits with same energy level (Fig. 8a), suggesting three *5d* orbits are necessary for hybridization because of their unoccupied states. When iodine and astatine form a two-dimensional structure, the *PDOS* of their *d* electron orbitals increase in comparison with the molecular state (Fig. S8, Fig. S9, Fig. S13- Fig. S14), indicating the crystal structure of iodine and astatine requires the participation of *d* electrons. As for fluorine molecules (Fig. S10), d orbital is absent. As for chlorine (Fig. S11), and bromine molecules (Fig. S12), the large energy difference between *d* and *s, p* orbital makes the *d* orbital difficult to participate in hybridization (Fig. S12). The iodine will connect with nearest other iodine atoms with hybridization orbits, forming σ bonds and β-iodiene. The *s* wave and $p_z$ wave (Fig. 7a, c) have much more contribution than $p_x$, $d_{xz}$, $d_{yz}$, $d_{zz}$ wave (Fig. 7b, d). The contributions in detail can refer to Fig. S8 and Fig. S9. An unpaired electron in $p_y$ wave has a possibility to form π band with contiguous atoms without contribution to hybridization orbits (Fig. 7).

For both the β-iodiene and B-astatiene, the $s$-$p^2$-$d^3$ hybridization orbits provide sufficient space for accommodating electrons, exhibiting energy overlapping among *s,p,d* wave (Fig. 8a).The valence electrons of β-iodiene and β-astatiene in molecular state are distributed in *s* orbital and *p* orbital, and a tiny fraction in the *d* orbital (Fig. S13, S14). Each atom of β-iodiene and β-astatiene bonds to six surrounding atoms (Fig. 8b red bonds), requiring six unpaired electrons. Thus, three paired electrons in the outermost shell of β-iodiene and β-astatiene jump to *d* orbital from *s* orbit and $p_x$ and $p_z$ orbits, which corresponds to the observed state density of d orbit in Fig. 7-Fig. 8.

As we all know, the unpaired electron in $p_y$ prefers to interact with its contiguous one to adjacent atoms, forming π bond, suggesting weak anti-localization property of electrons, and Dirac Fermions and topological non-trivial points can be survival.

β-astatiene and β-iodiene participate in orbital hybridization, belonging to sp2d3 hybridization within the same period.

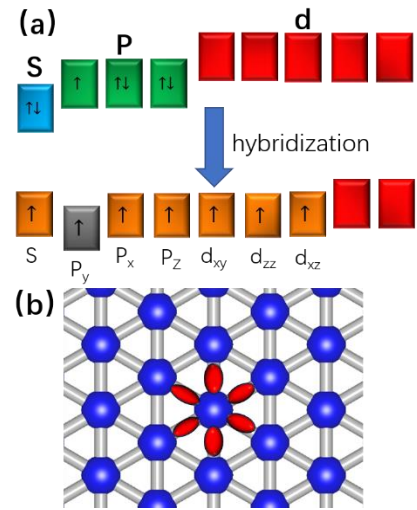

**Figure 8**(Color online) Hybridized orbital. (a): Hybridization of iodine and astatine atomic orbitals. The orange boxes are the ones that are participating in the hybridization. (b): Orbital profiles of β-iodiene and β-astatiene. The red ellipses are schematic of hybridized orbital.

Applying SOC can make the band gap reducing and band inversion even stronger than without (Fig. S15,S16). When no strain is applied, SOC effect drives valence bands away from each other and band gap reduction (Fig. 9a). When applying 9% compression strain, HOMO and LUMO are tangent when SOC effect is not taken into account, while the inversion occurs both in conduct band and valance band under SOC effect (Fig. b). When compression strain reaches 12.5% (Fig. 9c) and 15% (Fig.9d), the valence band inversion changes to G points instead of nearby pionts without SOC. The compression strain can creat band inversion but under different bands or orbit wavefunction.

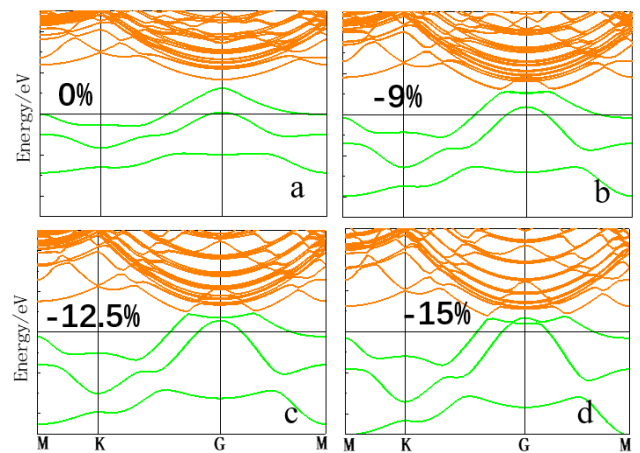

**Figure 9**(Color online) Band structure of β-astatiene with SOC, ($a_1$): without strain.($a_2$): with 9% compressive strain.($a_3$): with 12.5% compressive strain.($a_4$): with 15% compressive strain. The conduction band is in orange, and the valence band is in green.

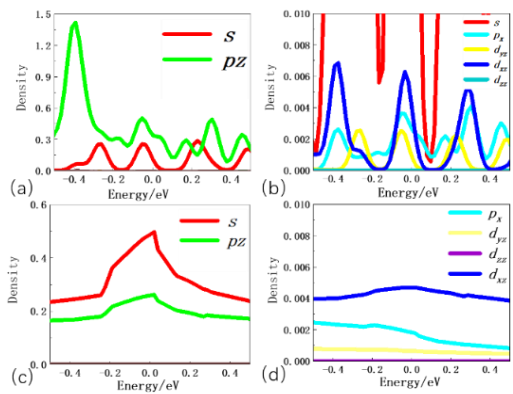

**Figure 7**(Color online) Partial density of states (PDOS) near the Fermi surface. (a):PDOS of β-astatiene. (b):Enlarged PDOS of β-astatiene. (c):PDOS of β-iodiene. (d):Enlarged PDOS of β-iodiene. The *s, $p_z$, $p_x$, $d_{xz}$, $d_{yz}$, $d_{zz}$* in

## Conclusion

New 2D materials: β-iodiene and β-astatiene can be stable and realized, if Iodine molecular crystal is considered, like graphite, natural iodine crystals are made of single layers of iodine that interact with each other by van der Waals forces[53]. The existence of VIIA elemental 2D configuration is reasonable and might be realized in experiment. The same period $sp^2d^3$ orbital hybridization permit the covalent to accommodate all the outmost electrons, and the high density of covalent bonds make the atoms tightly and steadily bounded into planar 2D configuration, displaying broadband gap and are insulators. Further applying strain can create their Dirac semi-metal and topological semi-metal properties. It is believed that by selecting the appropriate substrate, the method of CVD can be used to prepare iodide and astatine in the experiment.

## Experimental Section

First-principles calculations are performed by the plane wave code Vienna ab initio simulation package (VASP)[54], Generalized-gradient approximation (GGA) with the Perdew−Burke−Ernzerh of functional (PBE) was used to describe the exchange-correlation effects and electron-ion interactions[55]. The cutoff energy of the plane-wave expansion was set to 400 eV. All calculated equilibrium configurations are fully relaxed, the force and energy per atom are less than 0.01 eV/Å and $10^8$ eV respectively. Firstly, the lattice parameter of the system was optimized. Using 0.02 2π/Å accuracy for K point sampling. Gamma algorithm is used for all K point. A 30 Å vacuum layer is used to prevent the interaction of periodic structures. The phonon spectrum is calculated by DFPT algorithm with 2×2×1 supercell[56]. According to the formula:

$E_b = E_{at} - E_{sheet}$

(1)

the binding energy of materials ($E_b$) are calculated, where $E_{at}$ is the energy of an isolated spin polarized atom and $E_{sheet}$ is the energy of each atom in the two-dimensional materials[35].
By multiplying the two crystal axes of a and b by the same scaling coefficient, the biaxial strain of the crystal is achieved. Since the protocells a and b of P6/MMM space group are equivalent to each other, uniaxial strain is achieved by multiplying one of the crystal axes by a scaling factor and the other without any treatment.

## Acknowledgements

This work was supported partly by grants from the National Natural Science Foundation of China (No. 51472170 and 51771125). We also thank Dr. Ren for his help with VASP software.

**Keywords:** halogen • 2D materials • density functional calculations • electronic properties • Dirac point

# Supporting Information

# Stable halogen 2D Materials: the case of iodine and astatine

Xinyue Zhang[a], Yu Liu[b] and Qingsong Huang*[a]

[a] School of chemical engineering, Sichuan University, Chengdu, 610065, P.R.(China)

[b] Centre for Quantum Devices, University of Copenhagen & Microsoft Quantum Materials Lab Copenhagen

Denmark

## 1. Band structure of β-iodiene and β-astatiene under biaxial strain

The compressive strain can reduce the band gap, looking like the HOMO and LUMO are forced to close gradually. Once the HOMO contacts with LUMO, the band gap becomes zero, and π band survive. Further increasing compressive strain can make the contacted band inversion, mimicking the spring strips bends themselves upon compression force. The Dirac points and topological nontrivial points appear and become locked in the band structure.

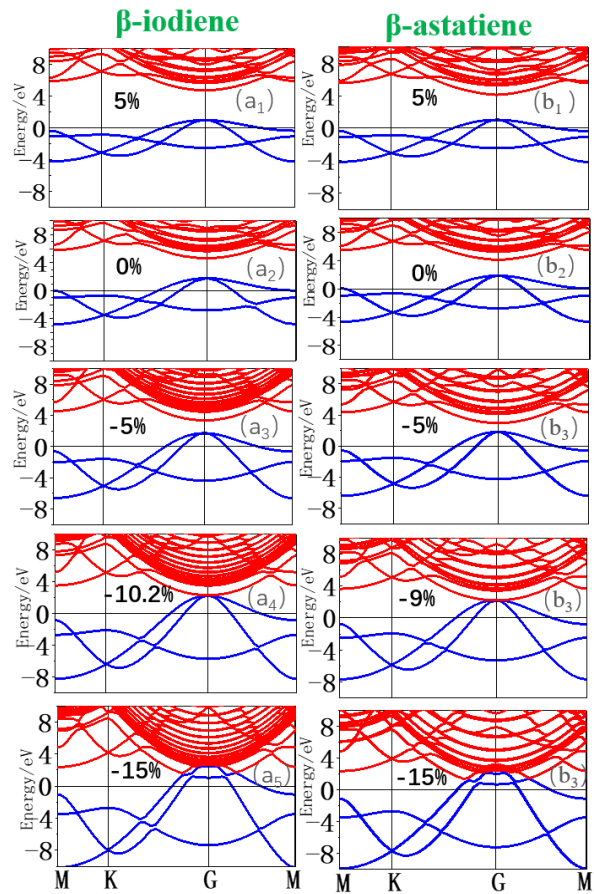

**Figure S1** Band structure of β-iodiene and β-astatiene under biaxial strain: (a1):β-iodiene under 5% tensile strain. (a2):β-iodiene without strain. (a3): β-iodiene under 5% compression strain. (a4):β-iodiene under 10.2% compression strain.(a5):β-iodiene under 15% compression strain. (b1):β-astatiene under 5% tensile strain. (b2):β-astatiene without strain. (b3):β-astatiene under 5% compression strain. (b4):β-astatiene under 9% compression strain.(b5):β-astatiene under 15% compression strain.

## 2. 3D band structure of β-iodiene under biaxial strain

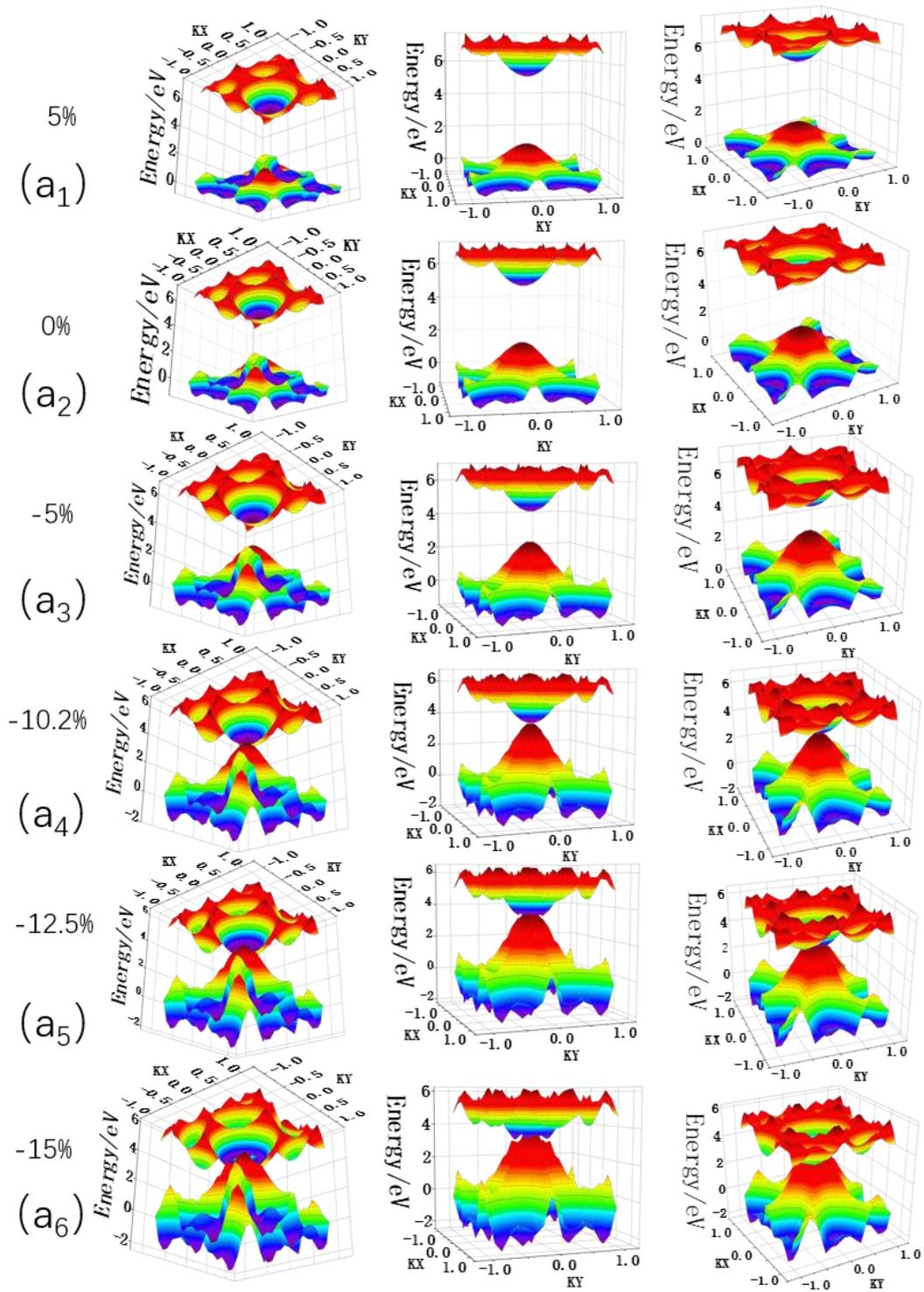

**Figure S2** 3D band structure of β-iodiene under biaxial strain in different visual angles. (a1):5% tensile strain. (a2):0% strain. (a3):5% compression strain. (a4):10.2% compression strain. (a5):12.5% compression strain. (a6):15% compression strain.

## 3. 3D band structure of β-astatiene under biaxial strain

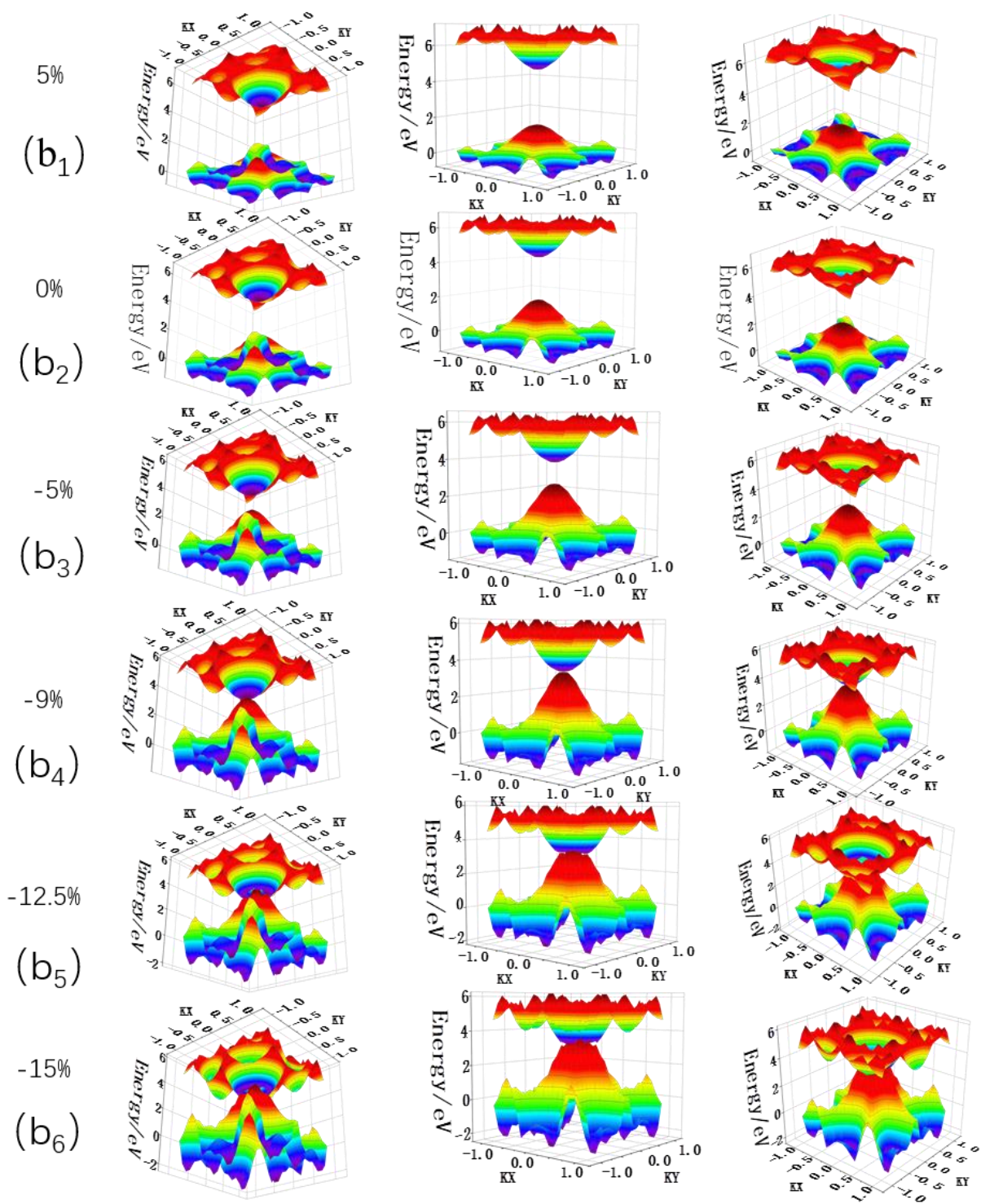

**Figure S3** 3D band structure of β-astatiene under biaxial strain in different visual angles. (b1):5% tensile strain. (b2):0% strain. (b3):5% compression strain. (b4):9% compression strain. (b5):12.5% compression strain. (b6):15% compression strain.

## 4. 3D band structure of β-iodiene under uniaxial strain

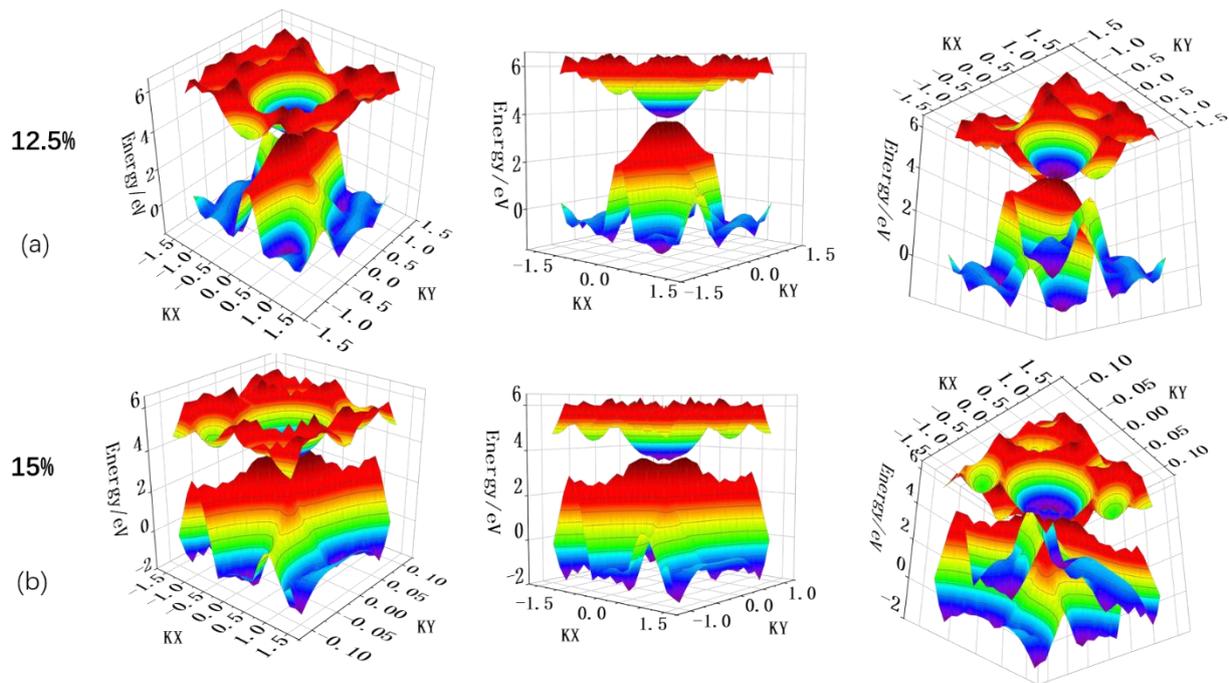

**Figure S4** 3D band structure of β-iodiene under uniaxial strain in different visual angles. (a):12.5%; (b):15%.

## 5. 3D band structure of β-astatiene under uniaxial strain

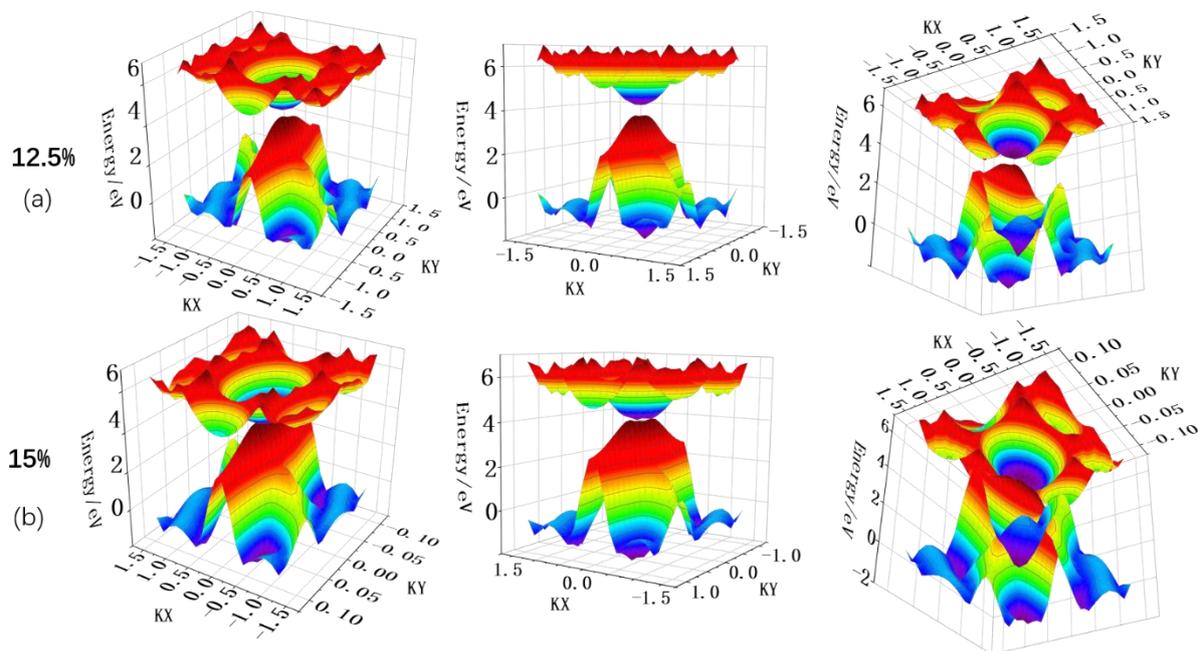

**Figure. S5** 3D band structure of β-astatiene under uniaxial strain in different visual angles. (a):12.5%. (b):15%.

## 6. β-iodiene and β-astatiene 3D band structure

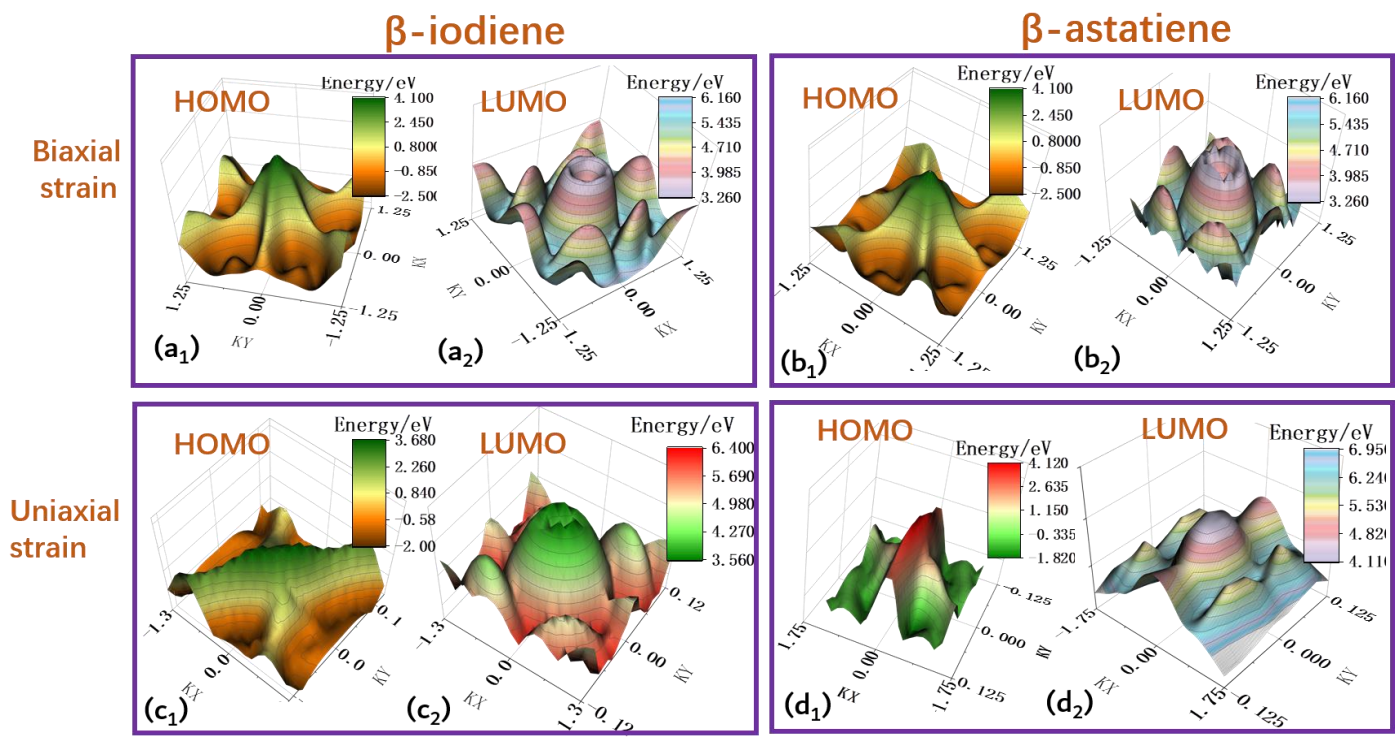

**Figure. S6** β-iodiene and β-astatiene 3D band structure and its constituent parts, including HOMO, LUMO, and related band inversion. ($a_1$-$a_2$): HOMO and LUMO 3D band structure of β-iodiene under 15% compression biaxial strain. ($b_1$-$b_2$): HOMO and LUMO 3D band structure of β-astatinene under 15% compression biaxial strain. ($c_1$-$c_2$): HOMO and LUMO 3D band structure of β-iodiene under 15% compression uniaxial strain. ($d_1$-$d_2$): HOMO and LUMO 3D band structure of β-astatinene under 15% compression uniaxial strain.

## 7. Structure and stability of hexagon iodine and astatine 2D structure

The virtual frequency in the phonon spectra suggests the hexagonal configuration is unstable for both iodine and astatine 2D structure. The existence of virtual frequency indicates that the hexagon β-iodiene and β-astatiene are unstable.

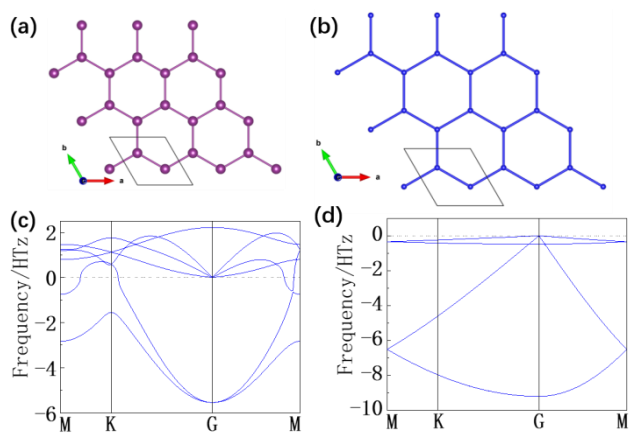

**Figure.S7** Structure and stability of hexagon iodiene and astatiene. Ball-and-stick model for (a): hexagonal β-iodiene model, and (b): hexagonal β-astatiene mode. Phonon spectrum for (c): hexagonal β-iodiene, and (d): hexagonal β-astatiene.

## 8. PDOS of β-iodiene

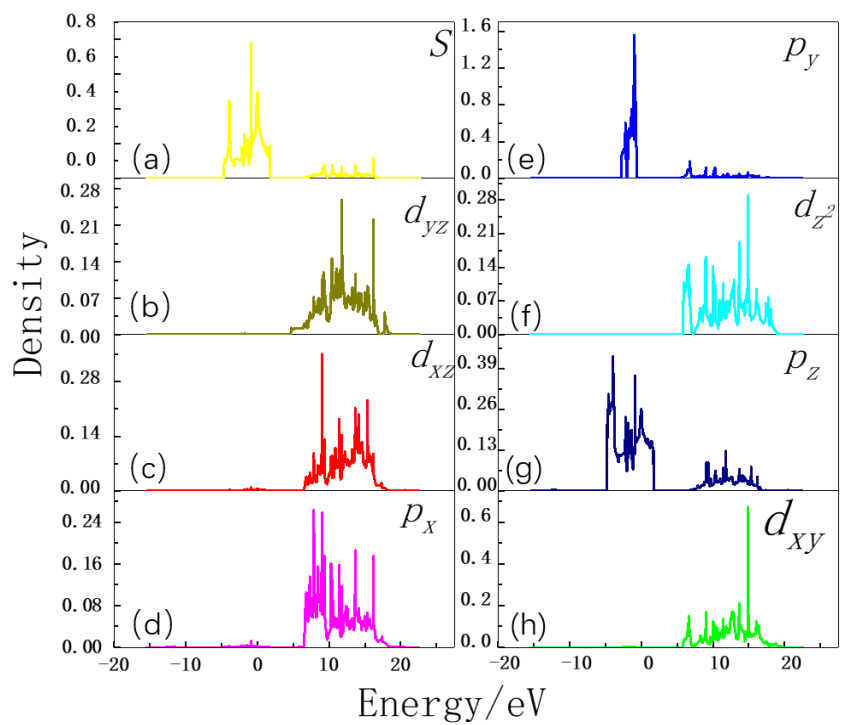

**Figure.S8** PDOS of β-iodiene,(a):$S$,(b):$d_{yz}$,(c):$d_{xz}$,(d):$p_x$,(e):$p_y$,(f):$d_{z2}$,(g):$p_z$,(h):$d_{xy}$.whose orbital energy of the d electron is over the Fermi surface, indicating that the d electron is in the same period as the s and p electrons.

## 9. PDOS of β-astatiene

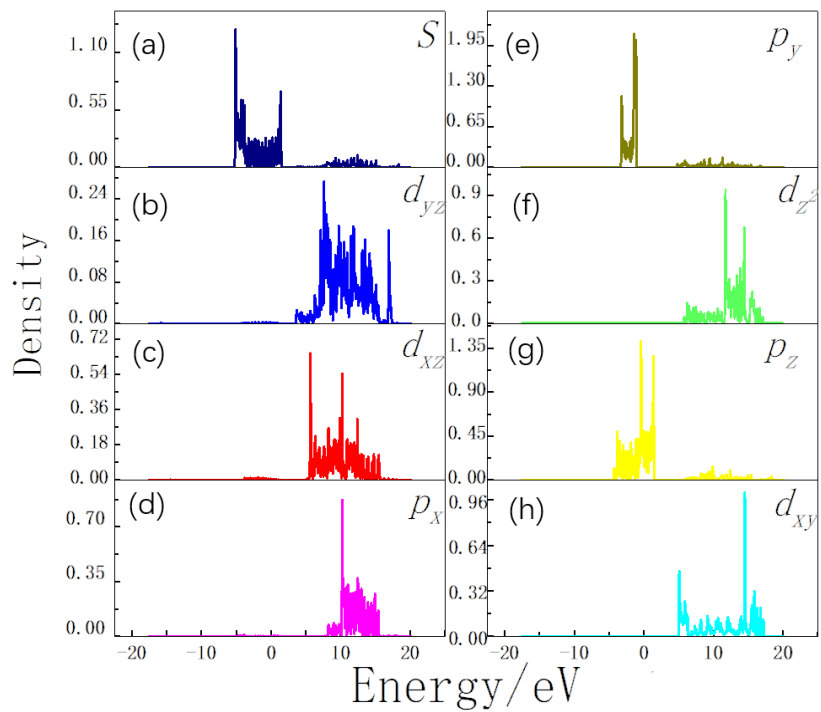

**Figure.S9** PDOS of β-astatiene.(a):$S$.(b):$d_{yz}$.(c):$d_{xz}$.(d):$p_x$.(e):$p_y$.(f):$d_{z2}$.(g):$p_z$.(h)$d_{xy}$.

## 10. PDOS of fluorine molecule

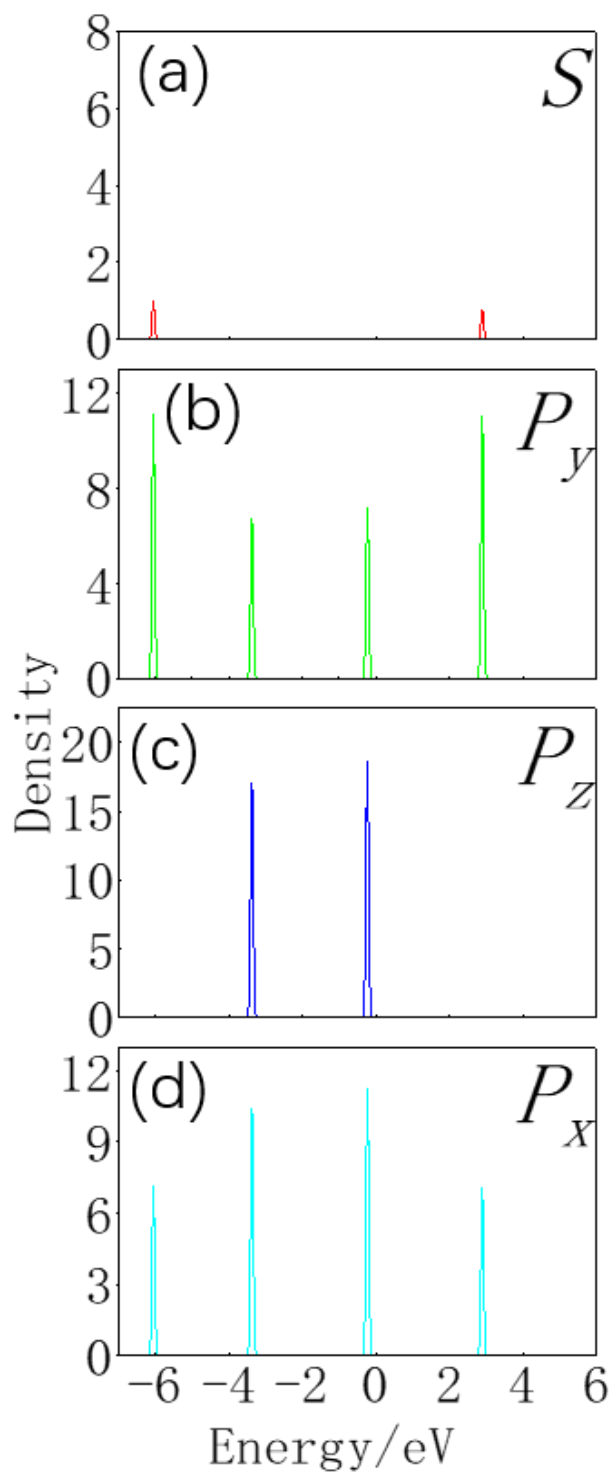

**Figure.S10** PDOS of fluorine molecule.(a):$S$.(b):$p_y$.(c):$p_z$.(d):$p_x$. Fluorine is in the second period, there is not $d$ orbital.

## 11. PDOS of chlorine molecule

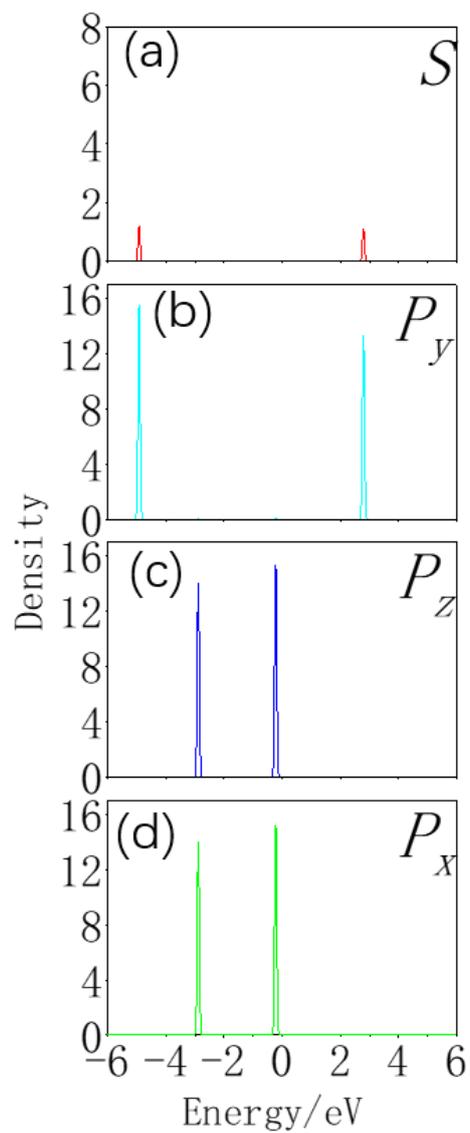

**Figure.S11** PDOS of chlorine molecule.(a):$S$.(b):$p_y$.(c):$p_z$.(d)$p_x$. Although chlorine has a d orbital in the third period, the PDOS of chlorine has no d orbital density.

## 12. PDOS of bromide molecule

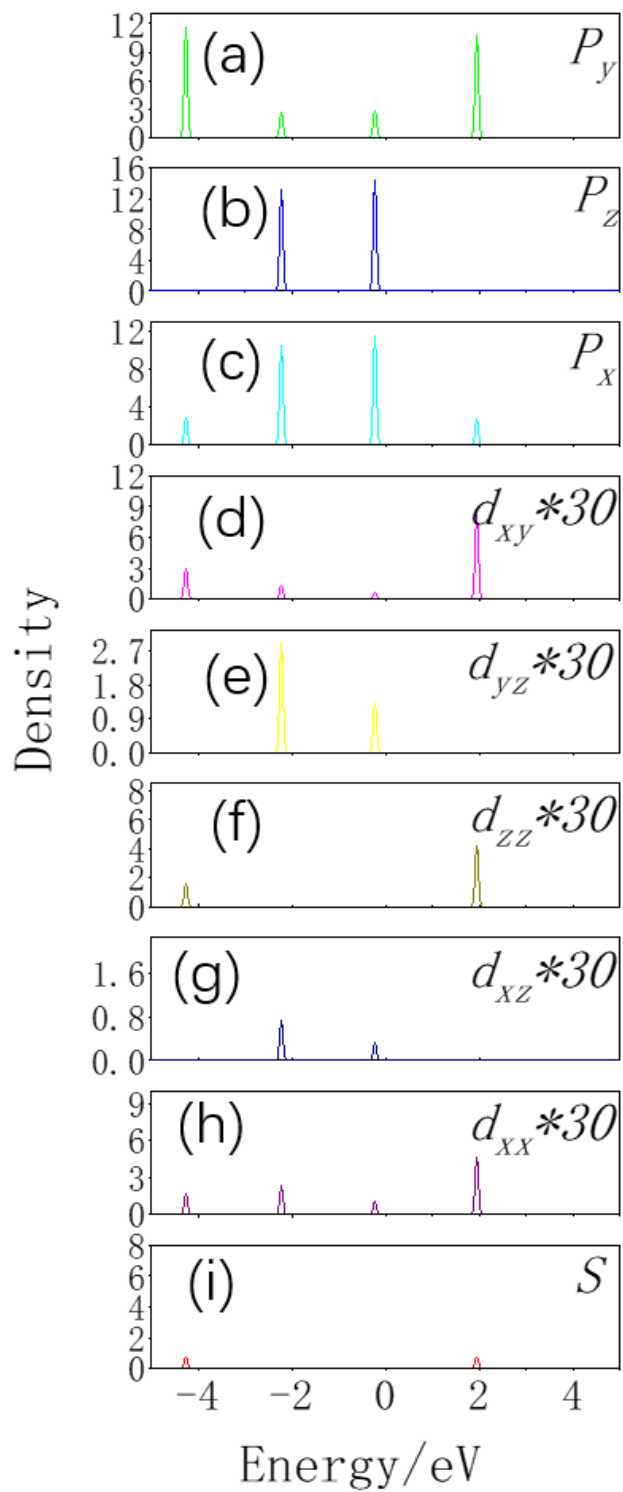

**Figure.S12** PDOS of bromide molecule.(a):$p_y$.(b):$p_z$.(c):$p_x$.(d):$d_{xy}$.(e):$d_{yz}$.(f):$d_{zz}$.(g):$d_{xz}$.(h):$d_{xx}$.(i):S. Bromine is in the fourth period, and *d* electron has a certain density of partial wave states.

## 13. PDOS of Iodine molecule

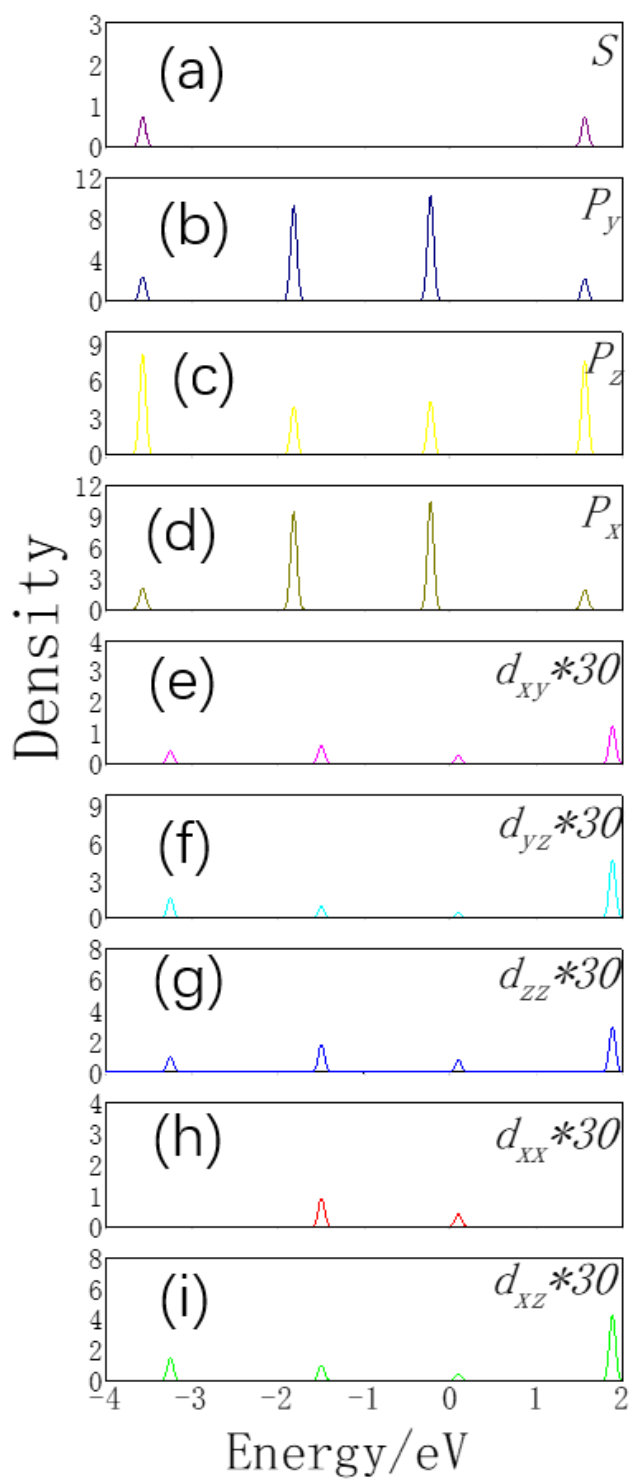

**Figure.S13** PDOS of Iodine molecule.(a):$S$.(b):$p_y$.(c):$p_z$.(d):$p_x$.(e):$d_{xy}$.(f):$d_{yz}$.(g):$d_{zz}$.(h):$d_{xx}$.(i):$d_{xz}$. With the increase of period number, due to the shielding effect of inner electrons, the energy range of s,p,d orbital decreases gradually, so that d electron can participate in orbital hybridization.

## 14. PDOS of astatine molecule

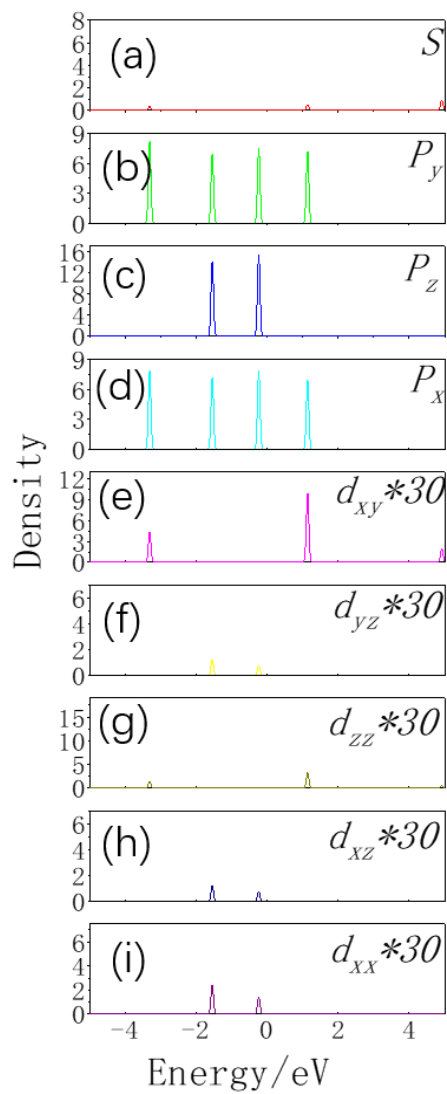

**Figure.S14** PDOS of astatine molecule.(a):*S*.(b):*p_y*.(c):*p_z*.(d):*p_x*.(e):*d_xy*.(f):*d_yz*.(g):*d_zz*.(h):*d_xz*.(i):*d_xx*.

## 15. Band structure of β-iodiene with SOC

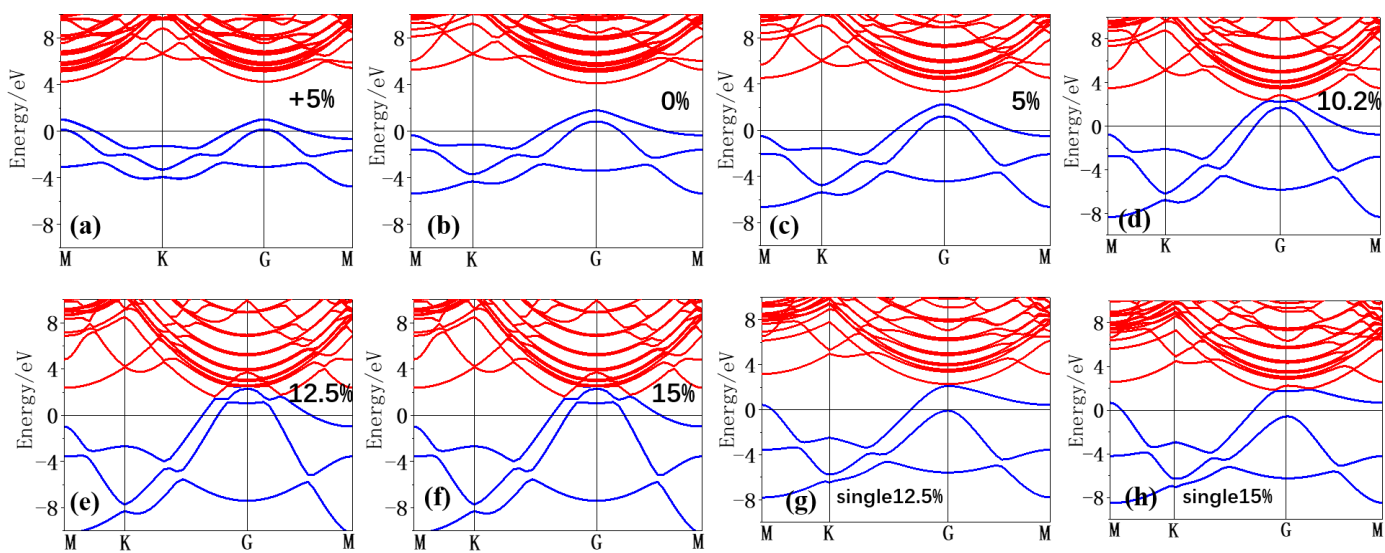

**Figure.S15** Band structure of β-iodiene with SOC. SOC band structure with (a):5% biaxial tensile strain. (b):no strain. (c):5% biaxial compressive strain. (d):10.2% biaxial compressive strain. (e):12.5% biaxial compressive strain. (f):15% biaxial compressive strain. (g):12.5% uniaxial compressive strain.(h): 15% uniaxial compressive strain.

## 16. Band structure of β-astatiene with SOC

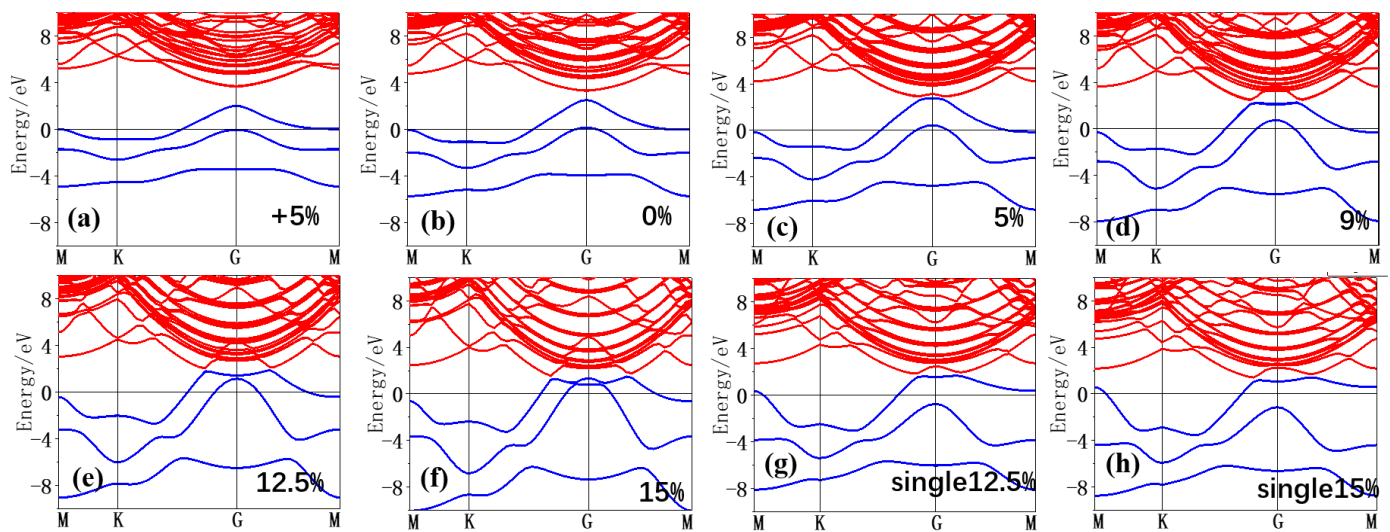

**Figure.S16** Band structure of β-astatiene with SOC. SOC band structure with (a):5% biaxial tensile strain. (b):no strain. (c):5% biaxial compressive strain. (d):10.2% biaxial compressive strain. (e):12.5% biaxial compressive strain. (f):15% biaxial compressive strain. (g):12.5% uniaxial compressive strain.(h): 15% uniaxial compressive strain.

## 17 Structural changes of buckling β-iodiene

Finally we explore buckling structure of β-iodiene and β-astatiene. As shown in the Fig. S17, buckling triangle β-iodiene became a bilayer flat structure after geometrical optimization.

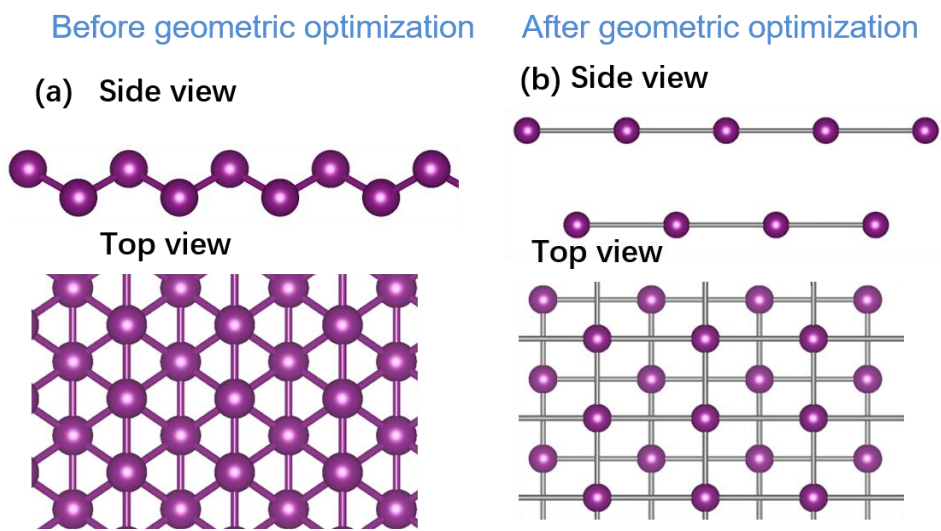

Figure.S17 Structural changes of buckling β-iodiene before and after geometric optimization. (a):Before geometric optimization.(b):After geometric optimization. Compared with figS16 a, b, buckling structure change to double layer plane structure.

## 18. Structural changes of buckling hexagon β-iodiene

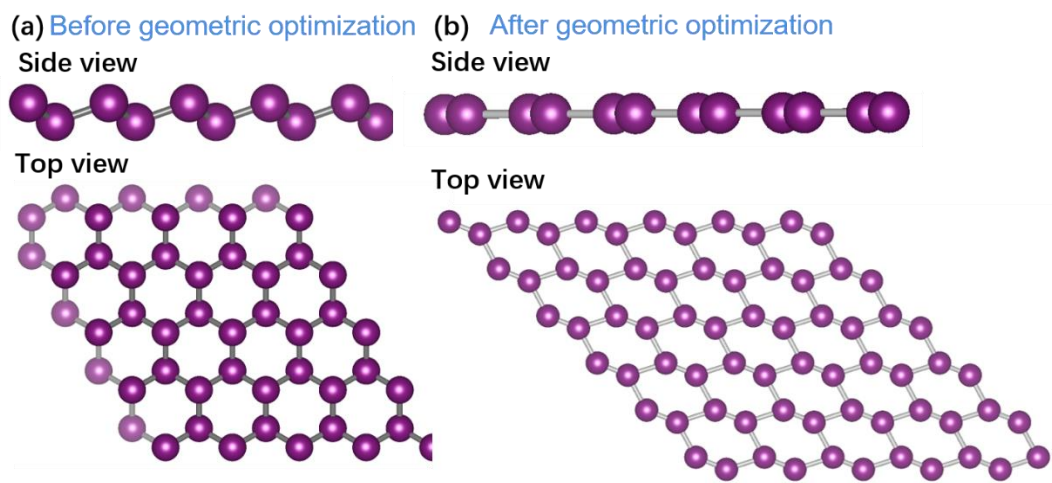

Figure.S18 Structural changes of buckling hexagon β-iodiene before and after geometric optimization. (a):Before geometric optimization. (b): After geometric optimization. Compared with Fig. S17 a, b, buckling structure change to plane structure.

## 19. Structural changes of buckling β-astatiene

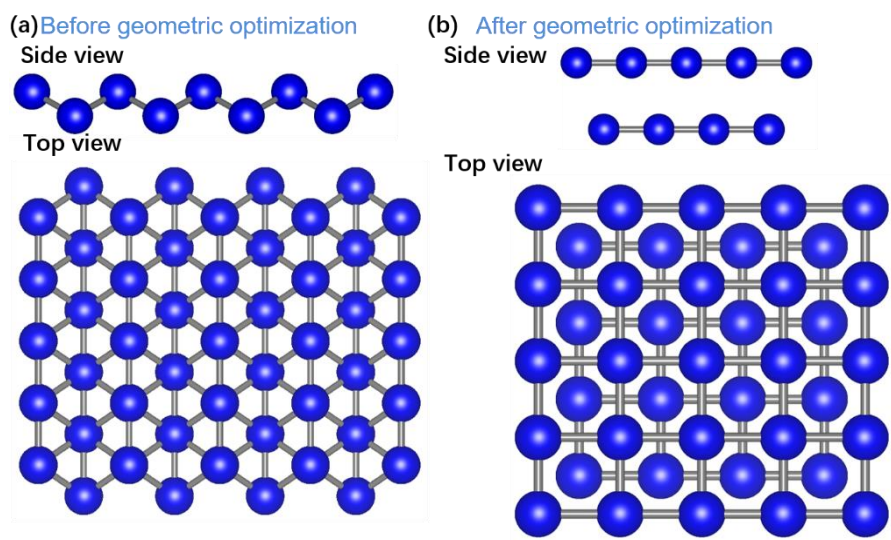

Figure.S19 Structural changes of buckling triangle β-astatiene before and after geometric optimization. (a):Before geometric optimization. (b):After geometric optimization.

## 20 Structural changes of buckling hexagon β-astatiene

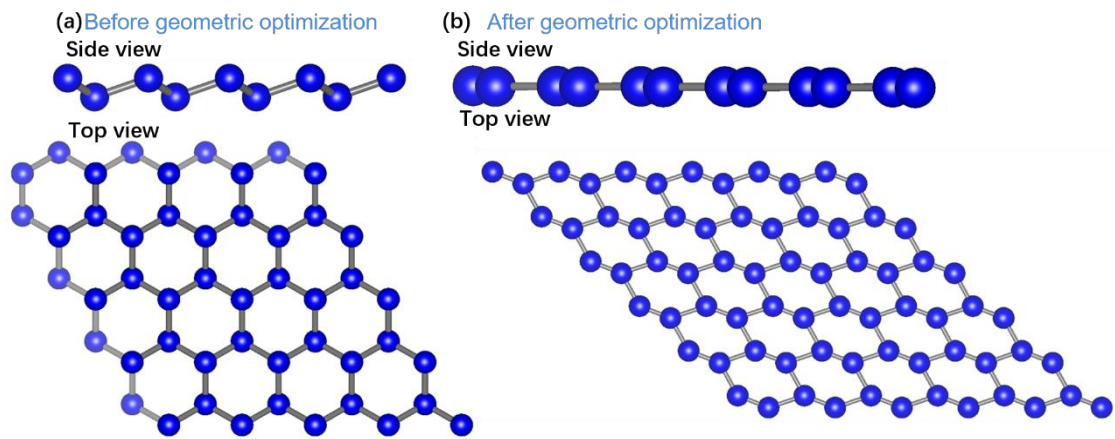

**Figure.S20** Structural changes of buckling hexagon β-astatiene before and after geometric optimization. (a):Before geometric optimization. (b):After geometric optimization.